\documentclass[aps,pre,showkeys,twocolumn,superscriptaddress]{revtex4-2}
\usepackage{amsmath}
\usepackage{amssymb}
\usepackage{epsf}
\usepackage{graphicx}
\usepackage{color}
\usepackage{newtxtext}

\begin{document}

\title{Comparison of Conformational Phase Behavior for Flexible and
    Semiflexible Polymers}

\author{Dilimulati Aierken}
\email{d.erkin@uga.edu}
\affiliation{Soft Matter Systems Research Group, Center for Simulational
    Physics, The University of Georgia, Athens, GA 30602, USA}
\author{Michael Bachmann}
\email{bachmann@smsyslab.org}
\homepage{http://www.smsyslab.org}
\affiliation{Soft Matter Systems Research Group, Center for Simulational
    Physics, The University of Georgia, Athens, GA 30602, USA}
\date{\today}

\begin{abstract}
    We employ the recently introduced generalized microcanonical
    inflection point method for the statistical analysis of phase transitions in
    flexible and semiflexible polymers and study the impact of the bending
    stiffness upon the character and order of transitions between random-coil,
    globules, and pseudocrystalline conformations. The high-accuracy estimates
    of the microcanonical entropy and its derivatives required for this study
    were obtained by extensive replica-exchange Monte Carlo simulations. We
    observe that the transition behavior into the compact phases changes
    qualitatively with increasing bending stiffness. Whereas the $\Theta$
    collapse transition is less affected, the first-order liquid-solid
    transition characteristic for flexible polymers ceases to exist once bending
    effects dominate over attractive monomer-monomer interactions.
\end{abstract}
%

\maketitle
%
\section{Introduction}
Statistical analysis methods have long been used for the identification,
characterization, and classification of phase transitions in complex systems.
Advanced computer simulation techniques, mostly based on Markov chain Monte
Carlo algorithms, helped advance the field beyond the deadlocks mathematical
approaches had run into. Nonetheless, the general idea remained the same.
Phase transitions were generally only considered in the thermodynamic limit,
which was historically introduced to make studies of macroscopic systems
mathematically tractable. In computational statistical physics, finite-size
scaling analysis provided the tool for extrapolating the results obtained in
simulations of systems of finite size toward this hypothetical limit. These
methods proved to be extremely successful, and for decades the scientific
community did not see any reason to change paradigms and consider a more
general perspective despite the fact that all systems in nature are finite
and systems on nanoscales moved into the focus of substantial interest in
microbiology and technology. Ignoring surface effects was also convenient.

Consequently, conventional canonical statistical analysis of phase
transitions rests on
the search for catastrophic behavior of thermodynamic state variables and
response functions in thermodynamic parameter spaces. Hence, significant
effort has been dedicated to the evaluation of critical exponents in
continuous phase transitions and the grouping of systems in universality
classes.

In recent years, however, the growing interest in systems on microscopic and
mesoscopic scales, which do not satisfy the criteria for the thermodynamic
limit, has shifted the perspective. Alternative approaches that enable
studies of phase transitions in systems of any size have turned out to
be promising. Modern interdisciplinary research of systems on
nanometer scales cannot ignore the fact that the surface of the system is at
least as important for the structural behavior of the system as the bulk
effects. For example, the functionality of biomacromolecules like proteins
depends on the folding process into stable geometric conformations. These
processes, which occur in a thermal environment, resemble
transitions between disordered and ordered phases known from macroscopic
systems. In fact, for very long polymers, these structure
formation processes \emph{are} phase transitions, even in the strict
conventional sense. However, many types of heterogeneous polymers like
proteins cannot be
scaled up, but still show clear transition features. This makes the extension
of the established theory of phase transitions a necessity.

However, canonical statistical methods create ambiguities, which render the
unique characterization of phase transitions problematic. A prominent example
is the peak in the specific heat curve of the one-dimensional Ising model. As
it is also finite in the thermodynamic limit, one does not associate a phase
transition with it. But how to interpret a peak in response quantities, if
the thermodynamic limit for this system does not exist? In recent years, it
has nonetheless become common to consider peaks or ``shoulders'' transition
signals in curves of response
quantities for finite systems. This appears to be
an inconsistent approach and renders, in fact, the choice of the temperature
as the basic thermodynamic state variable questionable.

Combining results of previous studies of the microcanonical Boltzmann
entropy~\cite{gross1} and the principle of  minimal
sensitivity~\cite{stevenson1,stevenson2}, we recently
developed a systematic approach to phase transitions in systems of any size
by identifying least-sensitive inflection points in the entropy and its
derivatives~\cite{qb1}. This analysis method even allows for the
classification of phase transitions in analogy to Ehrenfest's classification
scheme in thermodynamics, which is based on derivatives of thermodynamic
potentials~\cite{ehrenfest1}.

In this paper, we extend previous Monte Carlo computer simulation studies of
a coarse-grained model for flexible polymers~\cite{qlkpwb1} and analyze
structural transitions in semiflexible polymers by means of the generalized
microcanonical least-sensitive inflection point method. We thoroughly
investigate the changes of phase behavior of a coarse-grained model for
flexible and semiflexible polymers.

The paper is organized as follows. In Sec.~\ref{sec:mod}, we describe the
flexible and semiflexible variants of the coarse-grained polymer model used in
this study, as well as the numerical methods for the stochastic simulation of
the model and for the statistical analysis of the data obtained in the
simulations. The results of the canonical and microcanonical analysis
are
discussed and the phase behavior of the polymers is interpreted in
Sec.~\ref{sec:results}. This section also contains a detailed comparison of
geometric properties of the lowest-energy conformations found for each system.
Eventually, the paper is concluded by the summary in Sec.~\ref{sec:sum}.
%
\section{Models and Methods}
\label{sec:mod}
For this study we employed a generic
coarse-grained model for polymers in implicit solvent that allows us
to investigate and compare the phase behavior of flexible and semiflexible
polymer chains by means of advanced Monte Carlo computer simulation
methodologies.
Model, simulation methods, and the microcanonical inflection-point technique
we used for the statistical analysis of the conformational transitions are
described in the following.
\subsection{Generic model for flexible and semiflexible polymers}
The total energy of a polymer chain with $N$ monomers is composed of
contributions from bonded and non-bonded interactions between monomers. For
the interactions between non-bonded monomers, we employ the standard 12-6
Lennard-Jones (LJ) potential:
\begin{equation}
    V_{\mathrm{LJ}}(r)=4\epsilon_{\mathrm{LJ}}
    \left[\left(\dfrac{\sigma}{r}\right)^{12}-\left(\dfrac{\sigma}{r}\right)^{6}
        \right].
\end{equation}
Here, $r$ is the monomer-monomer distance, $\sigma=2^{-1/6}r_{0}$ is
the van der Waals distance associated with the
potential minimum at $r_0$, and
$\epsilon_{\mathrm{LJ}}$ is the energy scale. For computational efficiency,
we introduce a cutoff at $r_{c}=2.5\sigma$. Shifting the potential by the
constant $V_{\mathrm{shift}}=V_{\mathrm{LJ}}(r_c)$ avoids a discontinuity at
$r_c$. Thus the potential energy of non-bonded monomers is given by
\begin{equation}
    V_{\mathrm{NB}}(r)=
    \begin{cases}
        V_{\mathrm{LJ}}(r)-V_{\mathrm{shift}}, & r<r_{c},          \\
        0,                                     & \text{otherwise.}
    \end{cases}
\end{equation}
The elastic bond between two neighboring monomers is modeled by a potential
which combines a Lennard-Jones potential and the finitely extensible
nonlinear elastic (FENE) potential~\cite{Milchev2001,Kremer1990,Bird1987}:
\begin{equation}
    V_{\mathrm{B}}(r)=-\frac{1}{2}K_{\mathrm{f}}R^{2}\mathrm{ln}\left[1-\left(\frac{r-r_{0}}{R}
        \right)^{2}\right]+V_{\mathrm{LJ}}(r)-V_{\mathrm{shift}},
\end{equation}
where the FENE parameters are fixed to standard values $R=(3/7)r_{0}$ and
$K_\mathrm{f}=(98/5)\epsilon_{\mathrm{LJ}}/r_0^2$~\cite{Qi2014}. The bond length $r$ is restricted to
fluctuations within the range $[r_{0}-R, r_{0}+R]$. With these parameters the
minimum of $V_{\mathrm{B}}$ is located at $r_0$.

For semiflexible polymers, an additional potential related to the bending
stiffness is introduced. Any deviation from the reference angle $\theta _{0}$
between neighboring bonds is subject to an energy penalty of the form:
\begin{equation}
    V_{\mathrm{bend}}(\theta)=\kappa\left[1-\cos(\theta-\theta_{0})\right].
\end{equation}
If we represent a conformation of a polymer chain with $N$ monomers by
$\boldsymbol{X}=\left(\boldsymbol{r}_1,...,\boldsymbol{r}_N\right)$, where
$\boldsymbol{r}_i$ is the position vector of the $i$th monomer, the total
energy of the polymer is given by
\begin{equation}
    E(\boldsymbol{X})=\sum_{i>j+1}V_{\mathrm{NB}}(r_{i,j})+
    \sum_{i}V_{\mathrm{B}}(r_ {i,i+1})+\sum_{l}V_{\mathrm{bend}}(\theta_{l}),
\end{equation}
where $r_{i,j}=|\boldsymbol{r}_i-\boldsymbol{r}_j|$ is the distance between
monomers $i$ and $j$, and $\theta_{l}$ is the bond angle between two adjacent
bonds. The parameter $\kappa \geq 0$ controls the stiffness of the polymer
chain. For $\kappa=0$, the model describes flexible polymers.

In simulations and statistical analysis of the results, we set
the basic scales to the following values:
$k_\mathrm{B}=1$ (Boltzmann constant), $\epsilon_{\mathrm{LJ}} = 1$, and
$r_{0}=1$. For the reference bending angle, we chose $\theta_{0}=0$. The
flexible
chain with $N=55$ monomers has already been studied extensively in the past
and serves as the reference for the comparison with the semiflexible model.
This chain length is sufficiently short to recognize finite-size effects, but
long enough for the polymer to form a stable solid phase at low
temperatures~\cite{Schnabel2011}.
\subsection{Replica-Exchange Monte Carlo method and multiple-histogram
    reweighting}
For the simulations of the polymer, we employed the replica-exchange (parallel
tempering) Monte Carlo
method~\cite{Swendsen1986,Geyer1991,Huku1,Huku2,Earl2005}. Replicas of the
systems were simulated at different temperatures $T_{k} \in [0.1,5.0]$ with
$k=1,2,...,K$. The total number of temperature threads $K$ ranged from 40 to
50 in individual simulations. Up to ten independent runs were performed
for all $\kappa$ values studied, which allowed for the estimation of
numerical errors in all microcanonical quantities.

At each temperature $T_k$, Metropolis sampling was performed, which is based
on the acceptance probability:
\begin{equation}
    P(E_k,T_k;E'_k,T_k)  =
    \mathrm{min}\left\{e^{-(E'_k-E_k)/{k_{\mathrm{B}}T_k}}, 1\right\}.
\end{equation}
Here $k_{\mathrm{B}}T_k$ is the the thermal energy at the simulation
temperature $T_k$, $E'_k$ is the energy of the proposed state, and $E_k$
is the energy of the current conformation.

Local structural changes were achieved by displacement moves. In this Monte
Carlo update, a monomer $i$ is randomly picked and its position is modified by
a random shift $\Delta \boldsymbol{r}_{i}$ within a cubic box with edge lengths
$r_d$ surrounding the monomer. Before measurements can be performed, $r_d$ is
determined adaptively for each temperature thread to achieve a Metropolis
acceptance rate of about $50\%$.

In order to explore the conformation space more efficiently, rotational pivot
updates were also used. After 20 to 70 sweeps of displacement updates
(depending on the
temperature), we performed a sweep of pivot rotational updates. In a pivot
update, a monomer $i$ is randomly chosen as the pivot point and a
rotation axis vector $\boldsymbol{s}_i$. Then the section of the chain following
the pivot monomer was rotated about $\boldsymbol{s}_i$ by a randomly chosen
angle
taken from the interval $[0,2\pi]$.

Replicas were swapped between neighboring threads $k$ and $k+1$ after every
1500 sweeps with the exchange probability:
\begin{flalign*}
    & P(E_k,T_k;E_{k+1},T_{k+1})  = &
\end{flalign*}
\vspace*{-2.5em}
\begin{flalign}
    &&  \min\left\{\exp\left[\left(E_k-E_{k+1}
            \right)\left(\frac{1}{k_{\mathrm{B}}T_k}-\frac{1}{k_{\mathrm{B}}T_{k+1}}
            \right)\right], 1\right\},
\end{flalign}
where $E_k$ and $E_{k+1}$ are the energies of the replicas before the swap at
temperatures
$T_k$ and $T_{k+1}$, respectively.

In order to estimate the density of states of the system, we use the
multi-histogram reweighting method~\cite{Ferrenberg1988,Kumar1992}. For this
purpose, we measured the canonical histograms in each thread. By utilizing the
canonical distribution function $P_{\mathrm{can}}(E;T_k)\sim
    g(E)\exp\{-E/k_\mathrm{B}T_k\}$, estimates for the density of states can be
obtained by reweighting the energy histograms $h(E;T_k)$ measured in the
different simulation threads $k$,
$\overline{g}(E)=h(E;T_k)\exp\{E/k_\mathrm{B}T_k\}$. Using a
single histogram only covers a narrow range of energies effectively.
Combining the histograms obtained at different temperatures by employing the
multi-histogram reweighting method yields an estimator for
the density of states that covers the entire energy
range~\cite{Ferrenberg1988,Kumar1992}:
\begin{equation}\label{multi-1}
    \widehat{g}(E)=\frac{\sum_{k=1}^{K} h(E;T_{k})}{\sum_{k=1}^{K}
    M_{k}Z_{k}^{-1}e^{-E/k_{\mathrm{B}}T_k}}.
\end{equation}
Here $M_k$ is the total number of sweeps in the thread $k$ and $Z_k$ is an
estimator for the partition function:
\begin{equation}\label{multi-2}
    Z_k = \sum_{E} \widehat{g}(E)e^{-E/k_{\mathrm{B}}T_k}.
\end{equation}
Equations \eqref{multi-1} and \eqref{multi-2} are solved iteratively until
reasonable convergence has been achieved.
\subsection{Generalized microcanonical inflection-point analysis method}
Historically, phase transitions have been identified and classified by means
of discontinuities or divergences in thermodynamic state variables or response
functions. However, systems of finite size do not exhibit such obvious
signals, as these only occur in the thermodynamic limit. In this study, we use
the recently introduced generalized microcanonical inflection-point analysis
method~\cite{qb1} for the systematic identification and classification of
transitions in systems of any size. This method, which combines microcanonical
thermodynamics~\cite{gross1} and the principle of minimal
sensitivity~\cite{stevenson1,stevenson2}, has already led to novel insights
into the nature of phase transitions. Even the Ising model, which has been
excessively studied in almost a century, possesses a more complex phase
structure than previously known, as our most recent analysis
showed~\cite{Ked2020}.

Adopting Boltzmann's formula, the microcanonical entropy can be written as
\begin{equation}
    S(E)=k_{\mathrm{B}}\;\mathrm{ln}\;g(E),
\end{equation}
where $g(E)$ is the density of states with energy $E$. When not experiencing
phase transitions, the curves of entropy $S(E)$ and its derivatives exhibit
well-defined concave or convex monotony~\cite{qb1}. From canonical
statistical analysis
of first- and second-order transitions, it is known that entropy and/or
internal energy rapidly change, if the temperature is varied near the
transition point. This behavior corresponds to least-sensitive dependencies of
microcanonical quantities in the space of system energies. Thus, a phase
transition causes a least-sensitive inflection point in the entropy or its
higher derivatives and impacts their monotonic behavior. By systematically
analyzing these alterations, different types of transitions can be identified
and classified.

In this scheme, a first-order transition is signaled by a least-sensitive
inflection point with energy $E_{\mathrm{tr}}$ in $S(E)$. Therefore, the first
derivative, which is the inverse microcanonical temperature $\beta(E)$,
possesses
a positive valued minimum at $E_\mathrm{tr}$,
\begin{equation}
    \beta(E_\mathrm{tr})=\dfrac{\mathrm{d}
        S(E)}{\mathrm{d}E}\bigg|_{E=E_\mathrm{tr}}>0.
\end{equation}
Consequently, if there is a least-sensitive inflection point in $\beta(E)$,
the
phase transition is classified as a second-order transition. The derivative of
$\beta(E)$ has a negative-valued peak at the transition energy
$E_\mathrm{tr}$,
\begin{equation}
    \gamma(E_\mathrm{tr})=\dfrac{\mathrm{d}^2 S(E)}{\mathrm{d}
        E^2}\bigg|_{E=E_\mathrm{tr}}<0.
\end{equation}
More specifically, we call this an $\textit{independent}$ second-order
transition. This implies that there is another transition type:
$\textit{dependent}$ transitions. As the name suggests, dependent transitions
are associated with an independent transition of lower-order rank. It is
important to note that not all independent transitions have a dependent
companion. However, if it exists - and it can only exist at a higher
transition energy than its independent partner --- it servers as a precursor
of the independent transition in the disordered phase. For example, if this
inflection point is located within the convex region of the entropy, which is
associated with the first-order transition, the transition is a dependent
second-order transition. Consequently, there is a positive-valued minimum in
the derivative of $\beta(E)$ at the transition energy $E_\mathrm{tr}$,
\begin{equation}
    \gamma(E_\mathrm{tr})=\dfrac{\mathrm{d}^2
        S(E)}{\mathrm{d} E^2}\bigg|_{E=E_\mathrm{tr}}>0.
\end{equation}
Generally, for an independent transition of odd order $(2k-1)$ ($k$ is a
positive integer), the $(2k-1)$th derivative of $S(E)$ possesses a
positive-valued minimum,
\begin{equation}
    \dfrac{\mathrm{d}^{(2k-1)} S(E)}{\mathrm{d}
        E^{(2k-1)}}\bigg|_{E=E_\mathrm{tr}}>0
\end{equation}
and an independent transition of even order $2k$ is characterized by a
negative-valued peak in $2k$th derivative,
\begin{equation}
    \dfrac{\mathrm{d}^{2k} S(E)}{\mathrm{d} E^{2k}}\bigg|_{E=E_\mathrm{tr}}<0.
\end{equation}
A dependent transition of even order $2k$ is identified by a positive-valued
minimum in the $2k$th derivative,
\begin{equation}
    \dfrac{\mathrm{d}^{2k}
        S(E)}{\mathrm{d} E^{2k}}\bigg|_{E=E_\mathrm{tr}^\mathrm{dep}}>0,
\end{equation}
while for odd order $2k+1$:
\begin{equation}
    \dfrac{\mathrm{d}^{(2k+1)}
        S(E)}{\mathrm{d} E^{(2k+1)}}\bigg|_{E=E_\mathrm{tr}^\mathrm{dep}}<0.
\end{equation}
Dependent transitions are less common than independent transitions. In this
polymer study we did not identify any dependent transitions.

We used the B\'ezier method~\cite{Bachmann2014Book,Bezier1968,Gordon1974} to
smooth curves and calculate derivatives.
%
%
\section{Results}
\label{sec:results}
\subsection{Canonical analysis}
In the past, the common approach to the characterization of phase transitions
has been canonical statistical analysis. For infinitely large systems, one or
more derivatives of an appropriate thermodynamic potential such as the free
enthalpy or the free energy would exhibit nonanalytic behavior, if one of
the potential's natural variables was altered at the transition point.
Because it is easy to control in experiment, the analysis has typically been
done in the space of the canonical (or heat bath) temperature
$T_\mathrm{can}$. Discontinuities in the canonical entropy and specific heat
help identify and characterize discontinuous and continuous phase
transitions, respectively, in the thermodynamic limit. Later, this approach
was simply extended to systems of finite size that do not allow for the
hypothetical extrapolation toward the thermodynamic limit (for example,
heterogeneous polymers such as proteins).

However, transition features
in finite systems cannot be linked to nonanalyticities. Hence, \emph{peaks}
and \emph{shoulders} in response quantities like heat capacity or order
parameter fluctuations have usually been considered signals of
pseudotransitions in finite systems. The problem with this
approach is that extremal points in fluctuating quantities are not safe
indicators
of phase transitions. The probably most prominent example is the
specific-heat curve for the one-dimensional Ising model, which exhibits a
pronounced maximum at a certain temperature, but there is no obvious feature
that links this extremal thermal activity to a phase transition. In the
thermodynamic limit, this peak does not develop into a divergence.

Interestingly, if one plots the heat capacity for the 1D Ising system as a
function of the \emph{inverse temperature}
$\beta=1/T$, though, the peak vanishes and there is no
pseudotransition signal. Note that in the
microcanonical analysis the inverse temperature $\beta(E)=dS(E)/dE$ is the
more obvious thermodynamic variable than the temperature. Consequently,
microcanonical inflection-point
analysis does not show any transition signal for the 1D Ising system in equilibrium~\cite{Ked2020}.
\begin{figure}
    \centerline{
        \includegraphics[width=0.5\textwidth]{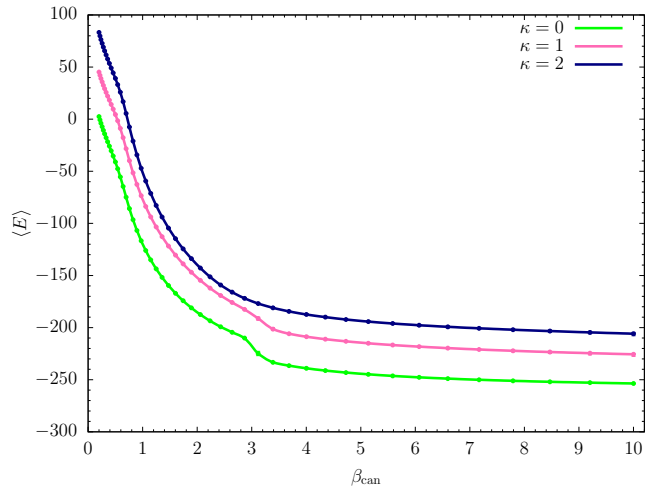}}
    \caption{Canonical mean energy $\langle E\rangle$ as a function of
        $\beta_\mathrm{can}$ for flexible ($\kappa=0$) and variants of semiflexible
        polymers with different bending stiffness ($\kappa=1,2$). Error bars are shown
        but smaller than the symbol sizes.}
    \label{fig:avg_energy}
\end{figure}

\begin{figure*}
    \centerline{
        \includegraphics[width=1.0\textwidth]{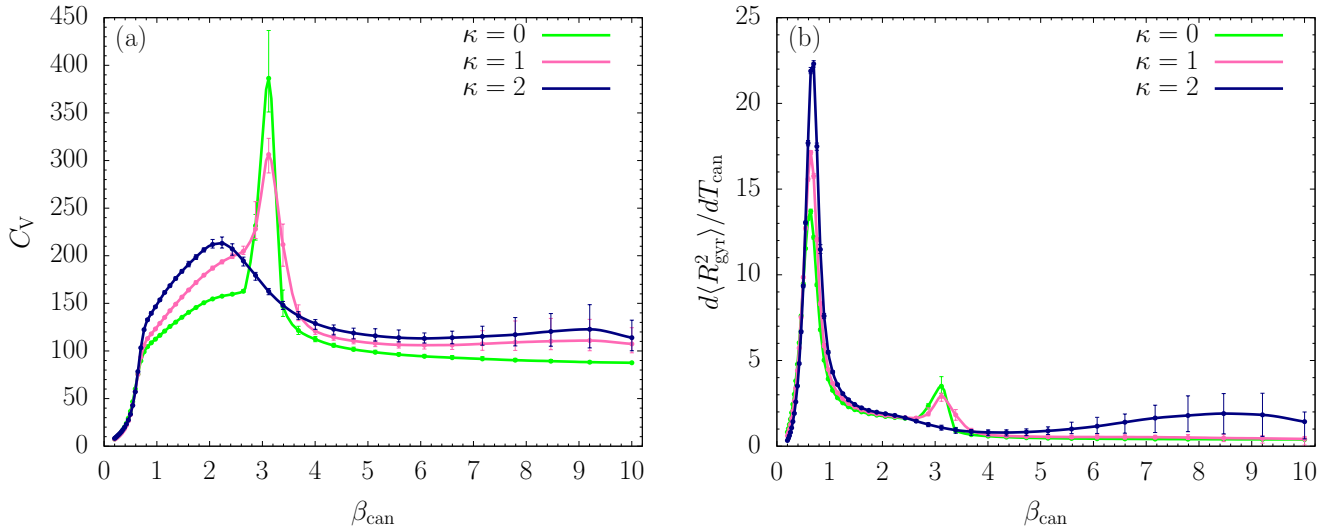}}
    \caption{Thermal fluctuations of (a) energy
        (heat capacity $C_V=d\langle E\rangle/dT_\mathrm{can}$) and (b) square radius
        of
        gyration ($d\langle R^2_\mathrm{gyr}\rangle/dT_\mathrm{can}$), plotted as
        functions of $\beta_\mathrm{can}$.}
    \label{fig:fluctuations}
\end{figure*}

In fact, historically the
(canonical) temperature was introduced when the need arose to assign a
reliable scale to the thermoscope to quantify the environmental heat
content, which ultimately lead Galilei to the introduction of the
thermometer at the end of the 16th century. It is ironic that the nowadays
most commonly used centigrade scale, the Celsius scale, was introduced in 1742
by Celsius in \emph{inverted} form, i.e., Celsius chose 0$^\circ$ as the
boiling point of water under norm conditions and 100$^\circ$ as the reference
point for the melting of ice. At about the same time, French scientist
Christin independently used a centigrade scale that
essentially inverted the original scale by Celsius, which is
what we nowadays refer to as the \emph{Celsius Scale}~\cite{bolton1}.

This brief historic recollection shows that the temperature and its scale
were introduced in an era, when it was still believed to be a
quantitative measure for heat, and the scale was ultimately linked to the
instrument and substances used for its measurement. Thus, there is no physical
reason that favors the historic quantity temperature over the inverse
temperature, which is the more natural variable in the microcanonical
analysis. To distinguish the microcanonical temperature from the heat bath
temperature, we denote the latter by $T_\mathrm{can}$ in the following
canonical analysis of structural transitions.

Consequently, we plot the canonical mean energy $\langle E\rangle$ as a
function of $\beta_\mathrm{can}=1/T_\mathrm{can}$ in Fig.~\ref{fig:avg_energy}
for the three models studied: $\kappa=0$ (flexible chain) and $\kappa=1,2$
(semiflexible variants). All curves show a sharp drop at
$\beta_\mathrm{can}\approx 0.7$. Around $\beta_\mathrm{can}\approx 3.2$, the
flexible chain ($\kappa=0$) experiences another significant drop in energy,
which is less pronounced for the semiflexible polymer with bending stiffness
$\kappa=1$. No obvious signal is visible on this level for the stiffer chain
with $\kappa=2$.

The plots of fluctuating quantities, such as the heat capacity  $C_V=d\langle
    E\rangle/dT_\mathrm{can}$ and the fluctuations of the mean square radius of
gyration, $d\langle R^2_\mathrm{gyr}\rangle/dT_\mathrm{can}$, allow for a more
detailed analysis. Both quantities are shown as functions of
$\beta_\mathrm{can}$ in Figs.~\ref{fig:fluctuations}(a)
and~\ref{fig:fluctuations}(b), respectively.
Whereas we certainly see changes in the curvature of the $C_V$ curves at
$\beta_\mathrm{can}\approx 0.7$, only the plots of the structural fluctuations
show sharp peaks.  These indicate the well-known $\Theta$ collapse transition
between extended, random-coil structures and compact globular conformations.
The transition signal is more pronounced for the stiffer chains. Since this
transition is more entropy than energy driven, it is not surprising that it
shows up more prominently in the structural rather than the energetic
fluctuations. This is different for the freezing transition at about
$\beta_\mathrm{can}\approx 3.2$, which is strongest for the flexible chain
($\kappa=0$). Large error bars at very small temperatures
($\beta_\mathrm{can}\approx 8.5$) prevent a conclusion about another separate
transition for the semiflexible polymer with $\kappa=2$, which, if it exists,
marks the transition into the global energy minimum basin.

\subsection{Microcanonical analysis of phase transitions}
\begin{figure*}
    \centerline{
        \includegraphics[width=0.8\textwidth]{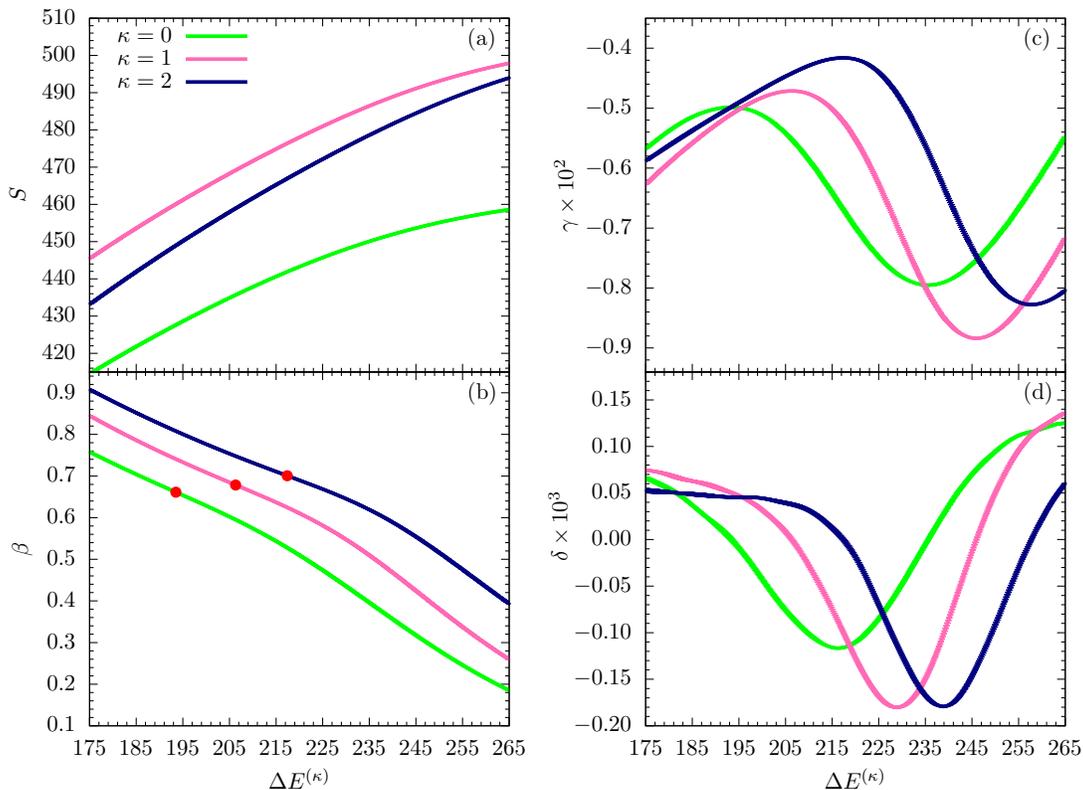}}
    \caption{(a) Microcanonical entropy $S$ and its derivatives
        (b) $\beta = dS/dE$, (c) $\gamma = d\beta/dE$, (d) $\delta = d\gamma/dE$ for
        the different models with $\kappa=0,1,2$ plotted as functions of the energy
        difference from the $\kappa$-dependent global energy minimum estimate. In this
        figure, we focus on the high-energy (or high-temperature) regime.
        Least-sensitive inflection points are marked by a dot.}
    \label{fig:micro_largeE}
\end{figure*}

\begin{figure*}
    \centerline{
        \includegraphics[width=0.8\textwidth]{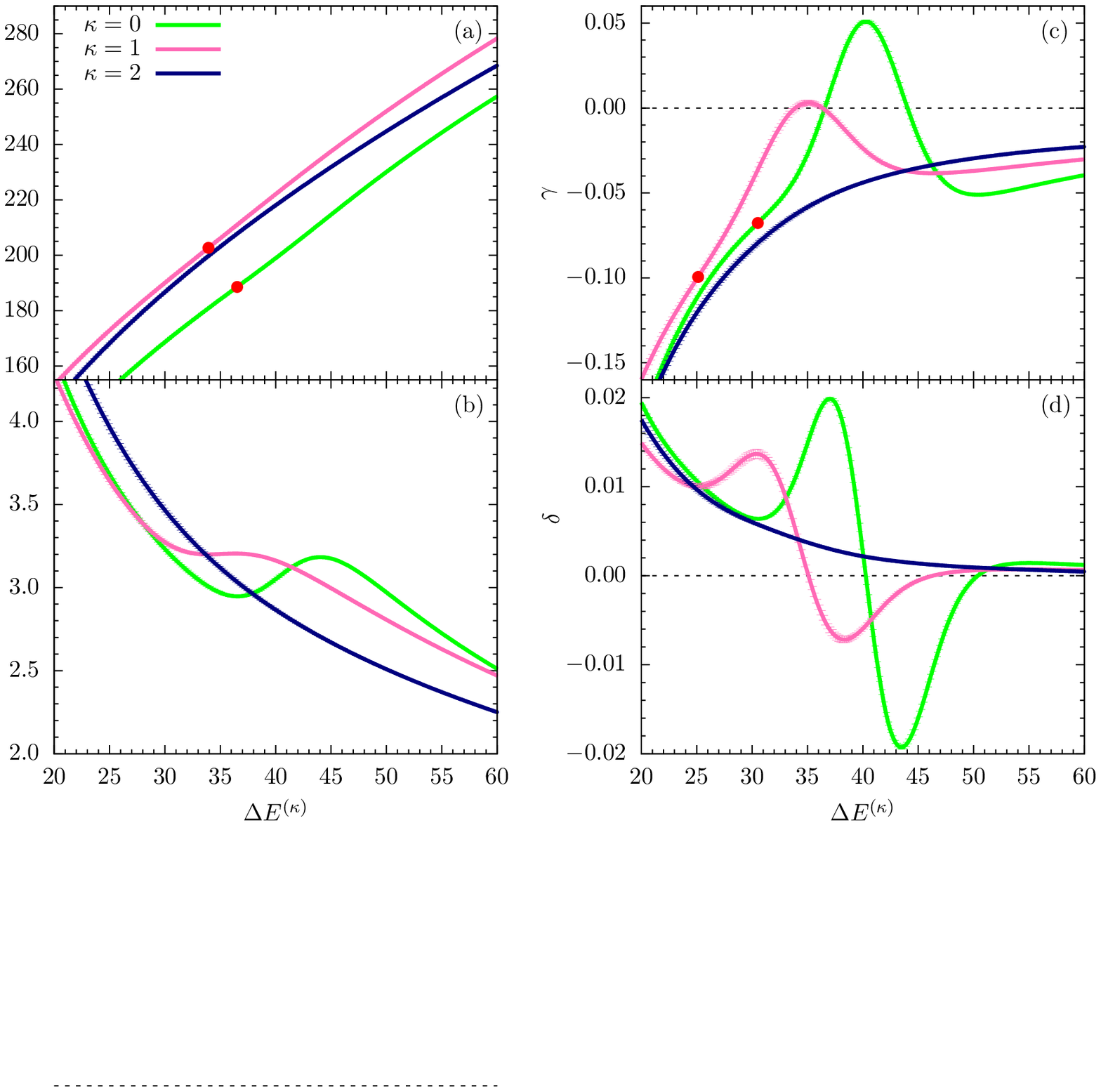}}
    \caption{Same quantities as in Fig.~\ref{fig:micro_largeE}, but plotted for
        an intermediate energy region.}
    \label{fig:micro_midE}
\end{figure*}

\begin{figure*}
    \centerline{
        \includegraphics[width=0.8\textwidth]{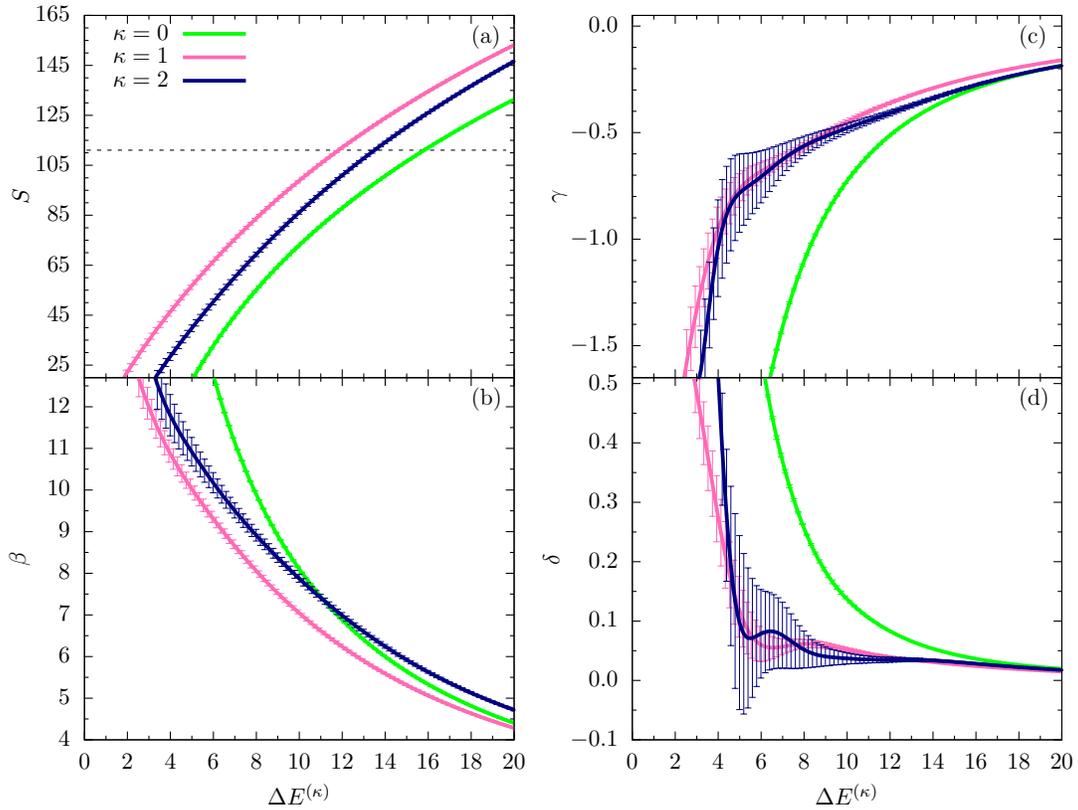}}
    \caption{Plots of the same microcanonical quantities as in
        Figs.~\ref{fig:micro_largeE} and~\ref{fig:micro_midE} in the lowest-energy
        regions of the three models.}
    \label{fig:micro_lowE}
\end{figure*}

In this section, we perform a full-scale microcanonical inflection-point
analysis of the three models that aims at the identification of all structural
transitions in these systems up to third order.
Figures~\ref{fig:micro_largeE},~\ref{fig:micro_midE}, and~\ref{fig:micro_lowE}
show the microcanonical entropy and its derivatives up to third order as
functions of the reduced energy
$\Delta E^{(\kappa)} = E-E_\mathrm{min}^{(\kappa)}$, i.e., we subtracted from
the energy the respective putative global energy minima
$E_\mathrm{min}^{(\kappa)}$ obtained for each system in the parallel tempering
simulations and verified by simulated annealing. This shift allows for an
easier comparison of the results.

As in the canonical analysis, we first discuss the low-$\beta$ (or
high-temperature) regime in the relevant energy space. The entropies plotted
in Fig.~\ref{fig:micro_largeE}(a) for all cases do not possess least-sensitive
inflection points and thus there is no first-order transition, as expected.
However, least-sensitive inflection points in the first derivative ($\beta$)
signal second-order phase transitions in all three systems, which reflect the
mostly entropic $\Theta$ collapse from random-coil to globular polymer
structures. The corresponding peak locations in the second entropy derivative
($\gamma$), shown in Fig.~~\ref{fig:micro_largeE}(c), allow for a unique
determination of the transition points in energy space and thus also in
$\beta$, thereby rendering the ambiguous canonical analysis of response
functions in the previous section obsolete. Since there are no least-sensitive
inflection points in the $\gamma$ plots, none of the systems undergoes a
third-order transition in this energy region.

Figure~\ref{fig:micro_midE} shows the same microcanonical quantities as
plotted in Fig.~\ref{fig:micro_largeE}, but for an intermediate energy range
that covers the inverse temperatures in the interval $\beta=(2,4)$. As
expected for flexible polymers, the entropy curve for $\kappa=0$ does exhibit
a least-sensitive inflection point, which corresponds to the minimum in the
backbending region found in the $\beta$ plot [Fig.~\ref{fig:micro_midE}(b)] at
about $\Delta E^{(\kappa)}=36.5$. The inverse temperature associated with it
is
approximately $\beta = 2.95$, which confirms earlier results for flexible
polymers~\cite{qlkpwb1}. The polymer undergoes a freezing transition from the
liquid-globular states into the solid phase, which for this model is dominated
by icosahedral structures. Remarkably, for the semiflexible polymers with
$\kappa=1,2$, the situation changes. For $\kappa=1$, there is still a weak
inflection point in the entropy, but the $\beta$ curve has already almost
plateaued at about the same energy value, where we found the first-order
transition for the flexible polymer. Consequently, the peak value of
the next derivative ($\gamma$) is virtually zero at this energy
    [Fig.~\ref{fig:micro_midE}(c)]. Therefore, the
freezing transition seems to turn from first order for the flexible polymer
($\kappa=0$) to second order for the semiflexible case with $\kappa=1$.
However, even more surprisingly, the transition behavior changes again for
the stiffer chain with $\kappa=2$: In fact, the transition signal has
completely vanished. The subsequent structural analysis of ground-state
properties will lend deeper insight into the reasons for these changes.

The second derivative $\gamma$ in Fig.~\ref{fig:micro_midE}(c) shows signs of
independent third-order transitions for $\kappa=0$ and $\kappa=1$ in the
ordered phase, but not for $\kappa=2$. These results suggest that freezing
into a unique and characteristic global energy minimum state is not possible
if bending effects overwhelm non-bonded interactions that are commonly
responsible for tertiary structure formation in polymeric systems. Local
bending effects turn into restraints that inhibit or at least suppress the
formation of symmetries. Previous studies have shown that bending restraints
play an important role in the formation of stable conformations in tertiary
assemblies of helical segments~\cite{wb1}.

%
%

\begin{figure*}
    \centerline{
        \includegraphics[width=0.8\textwidth]{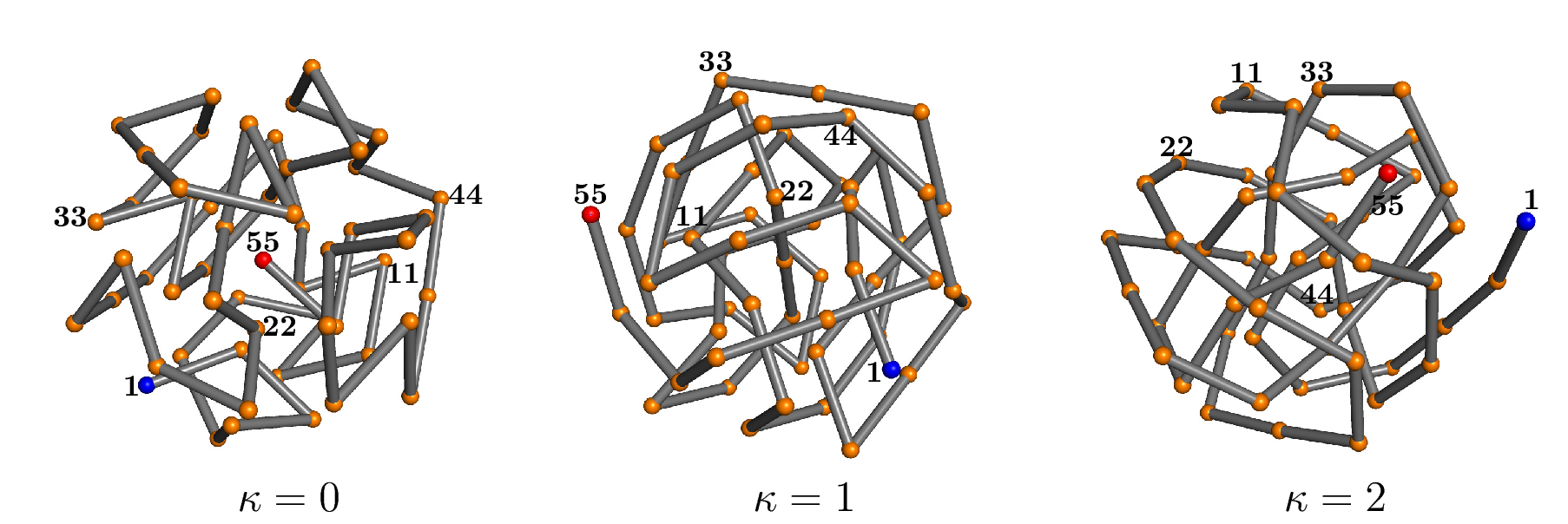}}
    \caption{Representatives of lowest-energy conformations for $\kappa=0,1,2$.
        Labels help to follow the chain from the first monomer (1, blue) to the last
        (55, red).}
    \label{fig:gs}
\end{figure*}

Eventually, the sequence of figures shown in Fig.~\ref{fig:micro_lowE} covers
the lowest-energy regime for all models compared in this study. In the
flexible case, no additional transitions are expected and, correspondingly, we
do not see signals in any of the derivatives up to third order that would
indicate a transition. The situation is potentially different for the systems
with bending restraint. Whereas we do not find indications of transition
signals in the entropy and inverse temperature curves for $\kappa=1$ and
$\kappa=2$ in Figs.~\ref{fig:micro_lowE}(a) and~\ref{fig:micro_lowE}(b),
respectively, the $\gamma$ results shown in Fig.~\ref{fig:micro_lowE}(c) might
tell a different story. The error bars are too large for an ultimate
conclusion, but it looks like a third-order transition develops close to the
ground state for the semiflexible polymer models.

The quantitative results of the microcanonical inflection-point analysis
obtained for the polymer systems studied here are listed in
Table~\ref{tab:transition}.
\begin{table}[t]
    \caption{Transitions found from microcanonical analysis for different
        flexible ($\kappa=0$) and semiflexible  ($\kappa=1,2$) polymers. Transition
        energy $E_{\mathrm{tr}}$, distance from putative ground-state energy $\Delta
            E^{(\kappa)}$, inverse microcanonical transition temperature
        $\beta_{\mathrm{tr}}$, and the order of the transition (classification) are
        listed.}
    \centerline{
        \begin{tabular}{ccccc}
            \toprule
            Bending stiffness     & $E_{\mathrm{tr}}$ & $\Delta E^{(\kappa)}$ &
            $\beta_{\mathrm{tr}}$ & Classification                                       \\
            \colrule
                                  & -231.2            & 30.5                  & 3.20 & 3 \\
            $\kappa=0$            & -225.2            & 36.5                  & 2.95 & 1 \\
                                  & -68.2             & 193.5                 & 0.66 & 2 \\
            \colrule
                                  & -205.8            & 25.1                  & 3.63 & 3 \\
            $\kappa=1$            & -197.0            & 33.9                  & 3.20 & 1 \\
                                  & -24.6             & 206.3                 & 0.68 & 2 \\
            \colrule
            $\kappa=2$            & 5.0               & 207.4                 & 0.70 & 2 \\
            \botrule
        \end{tabular}
        \label{tab:transition}
    }
\end{table}

\subsection{Structural analysis of lowest-energy conformations}

The lowest energies found in the simulations, which we consider best
estimates of the ground state energies, are $E_\mathrm{min}^{(0)}=-261.7$ for
the flexible polymer ($\kappa=0$), $E_\mathrm{min}^{(1)}=-230.9$ for the
semiflexible chain with $\kappa=1$, and $E_\mathrm{min}^{(2)}=-212.4$ if
$\kappa=2$. All values are given in units of the energy scale of the
Lennard-Jones potential, $\epsilon_\mathrm{LJ}$. Since the bonded interactions
(bond vibrations) are negligible near the ground state, the differences in the
ground-state energy estimates for the different models must be attributed to
the penalty paid for chain bending. The ground-state conformation for the
flexible polymer that does not experience these restraints is formed by
optimizing the non-bonded interaction and results, for this model, in an
ideally icosahedral structure. Turning on the bending restraint and setting
$\kappa=1$, the competition between nonlocal attraction and local repulsion
increases the tension in the icosahedral structure, although it still stays in
shape. For $\kappa=2$, however, it cannot maintain the optimal icosahedral
arrangement of monomers and rather forms an entangled structure of longer,
less bent, segments. As we have seen in the thermodynamic analysis of the
transitions, the phase behavior changes significantly with increasing bending
stiffness.
\begin{figure}[!htb]
    \centerline{
        \includegraphics[width=0.5\textwidth]{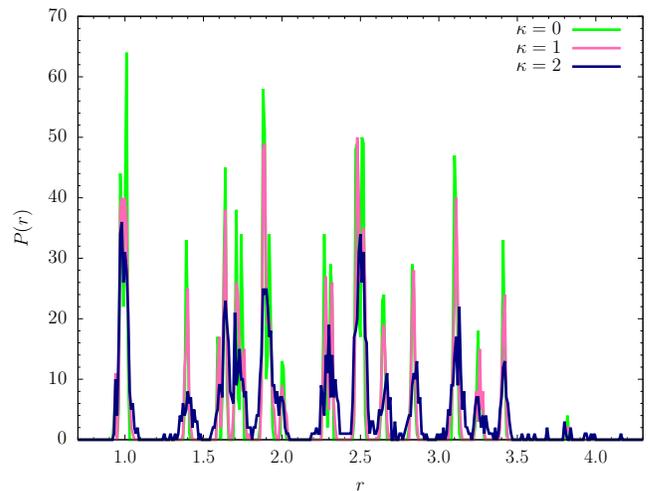}}
    \caption{Pair distribution functions for the lowest-energy states shown in
        Fig.~\ref{fig:gs} for $\kappa=0,1,2$.}
    \label{fig:pd}
\end{figure}

Representative lowest-energy conformations are shown in Fig.~\ref{fig:gs}.
For a more quantitative analysis let us take a look at the pair distribution
functions and the contact maps. We introduce the pair distribution function as
\begin{equation}
    P(r) = \sum\limits_{i<j} \Delta(r-r_{i,j}),
\end{equation}
where
\begin{equation}[h]
    \Delta(r-r_{i,j})=\left\{ \begin{array}{ll} 1, & |r-r_{i,j}|<r_t,    \\ 0,
                     & \mathrm{otherwise.}\end{array}\right.
\end{equation}
As a robust threshold for the necessary binning of the $r$ space, we chose
$r_t=0.01$. The histograms for the lowest-energy conformations shown in
Fig~\ref{fig:gs} are plotted in Fig.~\ref{fig:pd}. The perfect icosahedral
structure is only found for $\kappa=0$, whereas its decay is already visible
for the weaker semiflexible polymer ($\kappa=1$). The maximum number of
nearest-neighbor contacts ($r_{i,j}\approx 1$) found for $\kappa=0$ is not
reached in the semiflexible cases. The broadening of the peaks and additional
spikes not present for $\kappa=0,1$ are clear indicators that the ground-state
structure for the semiflexible polymer with $\kappa=2$ is not icosahedral.
Bending restraints prevent the formation of perfect symmetries and thus a
characterization of the $\kappa=2$ ground-state conformation as a distinct
crystalline or quasicrystalline structure is not possible. It rather resembles
tertiary folds of protein conformations, where effective bending restraints
and the local stable secondary segments purposefully prevent symmetric
arrangements of monomers. This enables different heteropolymers of similar
size to form distinct and functional individual conformations.
\begin{figure*}
    \centerline{
        \includegraphics[width=\textwidth]{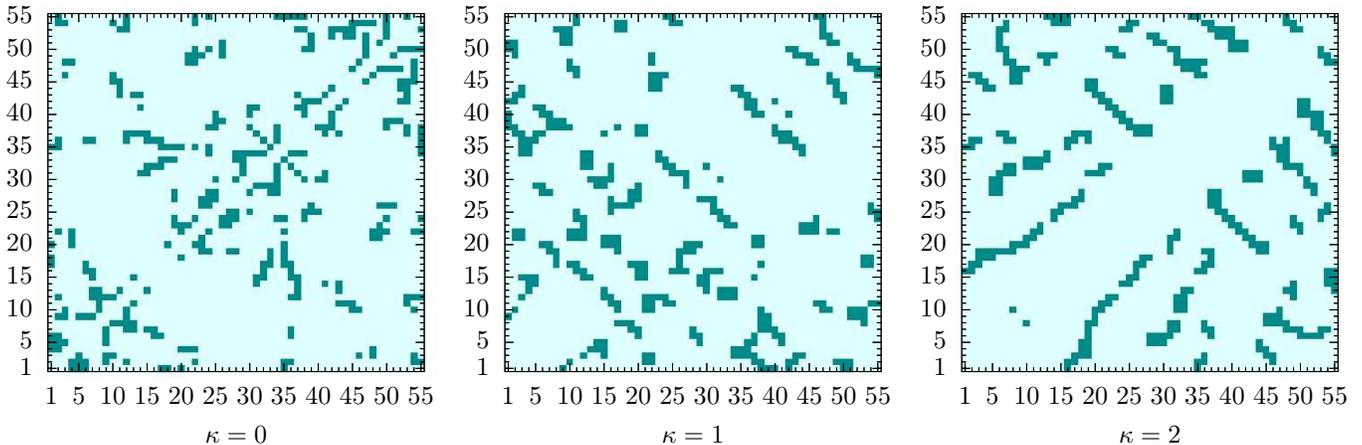}}
    \caption{Contact maps for the lowest-energy states shown in Fig.~\ref{fig:gs}
        for $\kappa=0,1,2$}
    \label{fig:cm}
\end{figure*}

This is obvious from the plots of the contact maps for the lowest-energy
states shown in Fig.~\ref{fig:cm}. In the grid spanned by the monomer labels,
we mark pairs of monomers with distances $r_{i,j}<1.2$. Bonded monomer pairs
are not included. For the flexible polymer, the contact map does not exhibit
particularly remarkable structural features. Since there are no bond angle
restraints, each monomer tries to maximize its number of nearest neighbors for
the energetic benefit. Icosahedral conformations with one of the end monomers
in the center are optimal. For the semiflexible polymer with $\kappa=1$, we
already see that changes occur. Monomers do not only try to maximize the
number of contacts, but three monomers connected by two adjacent bonds now
have to cooperate to minimize bending. Close to the diagonal, we find that
short anti-diagonal streaks form. These are clear indicators of turns with two
linear strands in contact with each other (hairpins). In protein folds, these
would be referred to as building blocks for $\beta$-sheets. For stronger
bending rigidity ($\kappa=2$), there are fewer, but longer such segments. We
also observe the formation of streaks parallel to the diagonal, which can be
associated with helical alignments. Both types of these secondary structure
elements can be seen in the geometric representation in Fig.~\ref{fig:gs}. The
hairpin-like sections are located in the interior and the helical segments
wrap around it. Of course, under these conditions, energetically favored
structures are not icosahedral. In fact, no global symmetries can be
identified in this tertiary fold. Although it is tempting to think of these
structures in analogy to protein folds, the lack of dihedral (torsion)
constraints, which separates secondary structure elements from each other in
the tertiary fold, prevents a more direct comparison.
%
\section{Summary}
\label{sec:sum}
The statistical analysis of structural transitions by the recently developed
generalized microcanonical inflection-point method~\cite{qb1} had already
yielded promising results in studies of flexible polymers~\cite{qlkpwb1}. For
this study, we extended the coarse-grained model by incorporating bending
restraints of different strengths to investigate the impact of bending
stiffness on the microcanonical entropy and its derivatives, which are used as
indicator functions for the systematic identification and classification of
phase transitions in systems of any size. In this coarse-grained model for
semiflexible polymers, the bending restraint is controlled by the bending
stiffness parameter $\kappa$.

Despite its simplicity, the model required a careful numerical treatment. For
this purpose, we performed extensive parallelized replica-exchange computer
simulations and verified the structural behavior in the lowest-energy basin by
means of stochastic methods optimized for this purpose such as simulated
annealing. Microcanonical thermodynamic properties were obtained from the
density of states, which is the main output of the multiple-histogram
reweighting method applied to the raw simulation data.

Based on the microcanonical results obtained in the simulations, we compared
the phase behavior of three model variants: purely flexible polymers
($\kappa=0$) and semiflexible polymers with $\kappa=1$ (identical energy
scales of attractive nonbonded monomer-monomer interactions and repulsive
bending strength) and $\kappa=2$ (bending restraints dominate over
monomer-monomer attraction). Our simulations reproduced previous results for
the flexible reference system very well, creating sufficient confidence for
the subsequent studies of the semiflexible polymers. Because of the additional
restraints, it was significantly more challenging to achieve in simulations of
semiflexible polymers the data quality necessary to enable an accurate
microcanonical analysis.

For the flexible polymer, we identified the known structural transitions,
i.e., the characteristic second-order transition associated with the $\Theta$
collapse from extended random coils to liquid globules and the first-order
liquid-solid transition, which is accompanied by an independent third-order
transition. The lowest-energy conformation is icosahedral, as expected. The
behavior does not change qualitatively for the weaker semiflexible case with
$\kappa=1$, where the effectively repulsive bending effects are competitive to
attractive monomer-monomer interaction. However, the larger number of possible
attractive monomer-monomer contacts represents an entropic advantage and thus
the icosahedral solid phase is still maintained. This is not the case anymore
if the chain is further stiffened by choosing $\kappa=2$. Whereas the $\Theta$
collapse remains widely unaffected, there is no obvious transition into the
solid phase anymore. Signals at very low energies at the third-order
transition level are too noisy to provide a clear picture. However, it is
obvious that the lowest-energy conformation found is a compromise of
satisfying both compactness and bending restraint. Longer, only slightly bent
segments form and assemble in a well-organized, coil-like structure. It is an
intriguing future task to extend our microcanonical study to stronger bending
restraints and to construct the complete hyperphase diagram in the spaces of
bending stiffness and inverse temperature and to apply it to other
modern applications in polymer science like hyperbranched
polymers~\cite{li1}.
%
%

%
\end{document}